\documentclass[twocolumn,pra,showpacs,preprintnumbers,amsmath,amssymb]{revtex4}

\usepackage{amsmath}
\usepackage{amssymb}
\usepackage{epsfig}

\begin{document}

\title{Optical fibers with interferometric path length stability by controlled heating for transmission of optical signals and as components in frequency standards}

\author{Holger M\"uller}
\affiliation{Physics Department, Stanford University, Stanford, CA 94305. Fax: (650) 723-9173, Email: holgerm@stanford.edu.}
\author{Achim Peters}
\affiliation{Institut f\"ur Physik, Humboldt-Universit\"at zu Berlin, Hausvogteiplatz 5-7, 10117 Berlin, Germany. Fax: +49 30 2093 4718, Email: achim.peters@physik.hu-berlin.de.}
\author{Claus Braxmaier}
\affiliation{Hochschule f\"{u}r Technik, Wirtschaft \& Gestaltung Konstanz,
Brauneggerstr.\,55, 78462 Konstanz. Fax: +49 (0)7531-206- 558, Email: braxm@fh-konstanz.de.}

\begin{abstract}
We present a simple method to stabilize the optical path length of an optical fiber to an accuracy of about 1/100 of the laser wavelength. We study the dynamic response of the path length to modulation of an electrically conductive heater layer of the fiber. The path length is measured against the laser wavelength by use of the Pound-Drever-Hall method; negative feedback is applied via the heater. We apply the method in the context of a cryogenic resonator frequency standard.
\end{abstract}

\maketitle
\section{Introduction}

The transmission of light through optical fibers is ubiquitous in optics. However, fibers are easily deformed. Vibrations and temperature changes thus cause an unwanted modulation of the optical path length $nL$ (where $n$ is the index of refraction of the fiber core and $L$ the mechanical length), thus phase modulating the transmitted signal. Even for fibers as short as a few meters under quiet laboratory conditions, the phase fluctuations easily exceed $2\pi$ on a timescale of seconds. When using fibers to transmit light between laboratories or different buildings, the corresponding broadening of the spectrum reaches several kHz \cite{Ma}. At first glance, this may appear negligible against the frequency of optical radiation. However, the tremendous progress in optical frequency stabilization has lead to lasers with a linewidth of one Hz and below \cite{Young,Eichenseer,MuellerMM,Webster,Notcutt}. Such lasers are applied, for example, in optical clocks \cite{opticalclock1,opticalclock2} and tests of fundamental laws of physics \cite{MuellerMM}. Precision atomic physics experiments use phase-locked lasers having  $\sim 10^{-4}$\,rad rms relative phase fluctuations \cite{MuellerPLL}. It is clear that transmission of such signals through fibers most significantly degrades their frequency and phase stability. 

Path length fluctuations may also have considerable effects in other fields: For example, quantum cryptography bridges distances exceeding 100\,km by fibers and is now beginning to be a commercial technology. Fiber length fluctuations are important in this context, since some concepts use interferometry to detect the quantum state of photons \cite{Gobby}. For global frequency comparisons at the targeted precision of next generation optical atomic clocks ($10^{-18}$), Doppler-shifts caused by the continental drift (at a velocity of $\sim 1$\,cm per year) will have to be controlled. Moreover, fiber length fluctuations have been identified as a limit on the frequency noise of optical comb generators \cite{Imai}. They also may play a role in space experiments like the LISA und LISA Pathfinder missions, which use Mach-Zehnder interferometers that have fibers in the interferometer arms. There is thus a strong and increasing need for a simple and reliable method to remove fiber path length fluctuations. 

\subsection{Previous approaches}

Removal of the unwanted effects of the path length fluctuations traces back to the `ultra-stable fiber optics link' over 29\,km used by Krisher {\em et al.} in a test of special relativity \cite{Krisher}. This is based on simultaneously sending two counterpropagating signals through the same fiber. By measuring the ratio of the sent and received frequencies at each side, the path length fluctuations are determined and can be corrected numerically. (Based on the same principle, frequencies can be accurately transferred to, from, and between spacecraft, with cancellation of the Doppler effect due to the relative motion. See, for example, \cite{Vessot}.) However, active cancellation of the noise generated by fibers has first been achieved by L.-S. Ma {\em et al.} \cite{Ma}. They use an acousto-optic modulator (AOM) to pre-modulate the phase of the light incident to the fiber with the negative image of the fiber noise, so that the noise is cancelled when the light passes the fiber. To gain the necessary information about the fiber induced noise, a ``probe" part of the signal is picked up at the fiber output and shifted by a constant frequency using a second AOM located at the output side of the fiber. It is then sent back through the fiber. Thus passing the fiber twice, the probe has picked up twice the phase shifts induced by the fiber. The beat note with the original light thus contains twice the fiber phase noise. The beat note frequency is electronically divided by two and used for driving the first AOM at the input side, which thus pre-modulates the light with the precise negative of the fiber phase noise. Restricting the bandwidth of the beat signal to about 10\,kHz (using a tracking VCO) avoids the introduction of additional wide-band noise into the signal path. A fiber with 25\,m length is used and phase noise is suppressed by 30\,dB to a level of -60\,dBc/Hz between 0.1 and 2\,kHz, corresponding to residual length fluctuations below $\sim 1\,$nm/$\sqrt{\rm Hz}$. Based on this method, a 3.5\,km fiber optics link has been established between the JILA and National Institute of Standards and Technology laboratories in Boulder, Colorado \cite{Ye2003}. The stabilization reduced the fiber-induced frequency noise at 1\,s integration time from about $2\times 10^{-14}$ to $3\times 10^{-15}$ (corresponding to $\sim 1\,$rad rms phase noise, or about 200\,nm rms length fluctuations); A 100\,m link between laboratories in the same building (the Max-Planck Institute for Quantum Optics in Garching, Germany) achieved a stability of 23\,mHz \cite{Eichenseer}. A 43\,km link reached $\sim 10^{-14}$ stability at 1\,s integration time \cite{Daussy}. Besides, this technique has applications in tests of fundamental physics \cite{Bize} (where a 180\,m long link is used), and may even be useful in quantum optics and atomic physics experiments, including quantum computing and cryptography \cite{Schmidtkahler}. 

\subsection{This approach}

In the approach described above, rather than stabilizing the path length, the resulting frequency fluctuations are corrected for by an AOM. While this is all that is needed in many cases, it has a couple of disadvantages: (i) The AOM has a finite deflection efficiency and thus reduces the transmitted power. (ii) When transmitting two laser frequencies $f_1$ and $f_2$, path length changes at a velocity $v$ produce two different frequency shifts of $f_1 v/c$ and $f_2 v/c$, respectively. Thus, the correction by the AOM works for one frequency only. This significantly reduces the degree of cancellation of noise if the frequencies $f_1$ and $f_2$ differ by more than a few per-cent. This may practically occur, for example when simultaneously transmitting a laser frequency and its second harmonic or even components of a femtosecond optical frequency comb. 

In this paper, we describe the use of controlled heating of (a part of) the fiber for stabilizing or modulating the optical path length. It achieves a performance similar to that of the method reported by Ma {\em et al.}. We directly stabilize the path length itself rather than compensating for the phase modulation caused by the fluctuations and thus avoid the above disadvantages.

This paper is organized as follows: In Sec. \ref{dynamic} we describe the dynamic response of the path length to the heater. The setup and its performance are described in section \ref{apparatus}. In Sec. \ref{coresec}, we show how the method was used in an experiment on frequency stabilization in a fiber-coupled cryogenic optica
l resonator (CORE).

\section{Dynamic response of the fiber}\label{dynamic}

Thermal time constants can easily be of the order of minutes or longer. Thus, if heating is to be used as an actuator in a fast servo loop that achieves sub-wavelength stability, the dynamic response of the path length to heating has to be considered carefully. Previous studies of the static \cite{Lagakos,Shibata} and the dynamic response \cite{Hughes,Schuetz} assume that the fiber surface is heated to a fixed temperature by an external thermal source. However, for fast temperature changes, this assumption becomes unrealistic, since the finite thermal conductivity of any material precludes the instantaneous application of a fixed temperature. Rather, the  dynamic characteristics of the thermal source itself have to be taken into account. We discuss the magnitude and the phase of the response, the latter being of prime interest for the design of a servo loop. 

In our model, the response of the path length $nL$ is given by changes in the index of refraction $n$ and the physical length $L$ of the fiber core:
\begin{equation}
\frac{\Delta(nL)}{nL}=\frac{\Delta n}{n}+\frac{\Delta L}{L}=\frac
1n \beta \Delta T(0)+\frac{\Delta L}{L}\,,
\end{equation}
where $\beta$ is the thermooptic coefficient and $\Delta T(0)$ denotes the temperature change of the fiber core. The simplest model of the fiber is that of a homogenous solid cylinder. The propagation of heat is governed by the diffusion equation
\begin{equation}
\dot T=\bar \lambda \triangle T+H\,,
\end{equation}
where $H(r,t)$ represents a heater which deposits an equal amount of thermal power into every volume element. $\bar\lambda=\lambda/(\rho c)$ is the thermal diffusion constant, given by the thermal conductivity $\lambda$, the mass density $\rho$, and the specific heat $c$. We first solve the homogenous equation with $H=0$. In cylindrical coordinates $r, \varphi, z$, specializing to the relevant case of no dependence on $\varphi$ and $z$, 
\begin{equation}
\dot T=\bar\lambda \left(\frac 1r \frac{\partial}{\partial
r}+\frac{\partial^2}{\partial r^2}\right)T\,.
\end{equation}
We separate $T(r,t)\equiv \mathcal T(t)\mathcal R(r)$ and obtain a radial and a temporal differential equation 
\begin{equation}
\mathcal R''+\frac 1r
\mathcal R'-\frac{i\omega}{\bar\lambda}\mathcal R=0\,,\quad \mathcal T'-i\omega \mathcal T=0\,.
\end{equation}
The separation constant is denoted $i\omega$. A solution to the temporal equation is proportional to $e^{i\omega t}$. The general solution to the radial equation can be written as
\begin{equation}
\mathcal R=AJ_0(\sqrt{i\omega/\bar\lambda}r)+BN_0(\sqrt{i\omega/\bar\lambda}r)\,.
\end{equation}
$J_0$ and $N_0$ are the Bessel and Neumann functions of zeroth order and $A$ and $B$ are linear coefficients. For the innermost layer of the fiber, the $B$ coefficient is required to vanishes since the Neumann function diverges for $r\rightarrow 0$.

Our model of the fiber consists of up to three concentric cylindrical layers (see Tables \ref{fibers} and \ref{matprop}): (i) the inner layer (radius $r_1$) represents both the fiber core and cladding. Although they  are doped to obtain slightly different index of refraction, the thermal and mechanical properties of the two are alike, so in the thermal model they can be considered as a single layer. (ii) A protective jacket of outer radius $r_2$, and (iii) a heater coating of outer radius $r_3$. (For the gold and aluminum coated fibers, the coating itself acts as the heater; therefore, $r_1=r_2$ and there are only two layers in the models for these fibers.) The linear coefficients $A_i$ and $B_i$ and the material parameters (Tab. \ref{matprop}) for each layer are denoted by subscripts $i$ that take the values 1,2,3. They are fixed by the boundary conditions that the temperature and the heat flow be smooth. 

\begin{table}
\centering
\caption{\label{fibers} Properties of fibers. See Tab. \ref{matprop} for material properties.}
\begin{tabular}{cccc}\hline
 & Plastic- & Aluminum- & Bare, gold- \\ 
 & jacketed & jacketed & coated \\ \hline
Layer 1 & Quartz glass & Quartz glass & Quartz glass \\
Layer 2 & Acryl &  &  \\
Layer 3 & Conductive varnish & Aluminum & Gold \\
$r_1$ [$\mu$m] & 60 & 60 & 60 \\
$r_2$ [$\mu$m] & 250 & & \\
$r_3$ [$\mu$m] & 400 & 250 & 70 \\ \hline
\end{tabular}
\end{table}

\begin{table}
\centering
\caption{\label{matprop} Material properties.}
\begin{tabular}{cccccc}\hline
& Acryl & Quartz & Al & Conductive & Au \\ & & glass & & varnish & \\ \hline
$\lambda$ [J/(K\,m\,s)]& 0.19 & 1.36 & 220 & 20 & 312 \\
$\rho$ [kg/m$^3$] & 2000 & 2200 & 2702 & $\sim 10^4$ & 19290 \\
E [GPa] & 3 & 75 & 71 & $\sim .005$ & 78 \\
$c$ [J/(kg\,K)] & 1700 & 729 & 896 & $\sim 400$ & 129 \\
$\alpha$ [$10^{-6}$/K] & 80 & 0.45 & 23.8 & $\sim 100$ & 14.3 \\
$\beta$ [$10^{-6}$/K] &  & 9 & &  \\ \hline
\end{tabular}
\end{table}

Within the heater volume, we assume a thermal source that is periodic in time: 
\begin{equation}
H=\frac{h}{\rho c}e^{i\omega t}\,,
\end{equation}
where $h$ represents the power density of the heater in W/m$^3$. The solution of the inhomogenous differential equation is the sum of a particular solution and the general solution of the homogenous equation. A particular solution 
\begin{equation}
T(r,t)=\frac{h}{i\omega\rho c}e^{i\omega t}
\end{equation}
can easily be found. We introduce the short notation 
\begin{equation}
Z_n(i,j)\equiv Z_n\left(\sqrt{\frac{i\omega}{\bar\lambda_i}}r_j\right)\,,
\end{equation}
where $Z_n$ denotes any of the functions $J_n$ and $N_n$. The boundary conditions are explicitly
\begin{eqnarray}
A_1 J_0(1,1)&=&A_2J_0(2,1)+B_2N_0(2,1)\,,\nonumber \\
A_1J'_0(1,1)&=&(\lambda_2/\lambda_1)[A_2J'_0(2,1) \nonumber \\ & & +B_2N'_0(2,1)]\,,\nonumber \\
A_2J_0(2,2)+B_2N_0(2,2)&=&A_3J_0(3,2)+B_3N_0(3,2)\nonumber\\ && +\frac{h}{i\omega\rho_3 c_3}e^{i\omega t}\,,\nonumber\\
A_2J'_0(2,2)+B_2N'_0(2,2)]&=&(\lambda_3/\lambda_2)[A_3J'_0(3,2)\nonumber \\ & & +B_3N'_0(3,2)]\,,\nonumber\\
A_3J'_0(3,3)+B_3N'_0(3,3)&=&0\,.
\end{eqnarray}
The last equation means that for now we assume no heat dissipation to the environment, the effects of which will be discussed later. This system of linear algebraic equations can be solved for the coefficients $A_i, B_i$. Since the solution is elementary but lengthy, we will not spell it out here. The temperature of the core and thus the associated path length change $\beta\Delta T(0)$ are given by $\Delta T(0)=A_1J_0(1,1)$. 

The additional strain term is caused by thermal expansion, which leads to axial strain (i.e., length change) \cite{Hughes,Schuetz}. This changes the path length $nL$ directly. (The temperature changes also cause radial strain which leads to a change in the phase velocity of light in the waveguide \cite{Hughes,Schuetz}. However, as this effect contributes only about 10\% of the total effect, we can neglect it here.) Each infinitesimal layer of the fiber wants to adjust its length according to the local temperature. For alternating temperatures, the temperature distribution is inhomogenous; stress builds up, which propagates to other layers at the speed of sound. For the time-scales relevant here, we can assume instantaneous sound propagation. Hence, the length of the fiber is then given by a balance of forces 
\begin{equation}
\int r\alpha(r)E(r)T(r)dr=\Delta L \int r\lambda(r)dr\,.
\end{equation}
Here, $\alpha$ is the thermal expansion coefficient and $E$ the bulk modulus. These quantities appear under the integrals, because they are different for the layers of the fiber. Hence,
\begin{equation}
\Delta L=\frac{\int \alpha(r)\lambda(r) T(r) r dr}{\int r\lambda(r) dr}\,.
\end{equation}
Using $\int x Z_0(x)dx=xZ_1(x)$ and defining $r_0\equiv0$,
\begin{eqnarray}
\Delta L(\omega)= \left(2\sum_{k=1}^3 E_k(r_k-r_{k-1})^2\right)^{-1}\nonumber \\ \times \left(\frac{ h}{i\omega\rho_3c_3}\frac{r_3^2-r_2^2}{2} +\sum_{k=1}^3 \frac{E_k\alpha_k\sqrt{\bar\lambda_k}}{\sqrt{i\omega}} \right. \nonumber \\ \left. \times \left[A_krJ_1\left(\sqrt{\frac{i\omega}{\bar\lambda_k}}r\right)+B_krN_1\left(\sqrt{\frac{i\omega}{\bar\lambda_k}}r\right)\right]^{r_k}_{r_{k-1}}\right)\,.
\end{eqnarray}
The core temperature and the strain term have the same sign for slow temperature variations. In what follows, we study the path length change as given by these terms for the fibers listed in Tab. \ref{fibers}.

\paragraph{Plastic-jacketed fiber}

\begin{figure*}
\centering
\epsfig{file=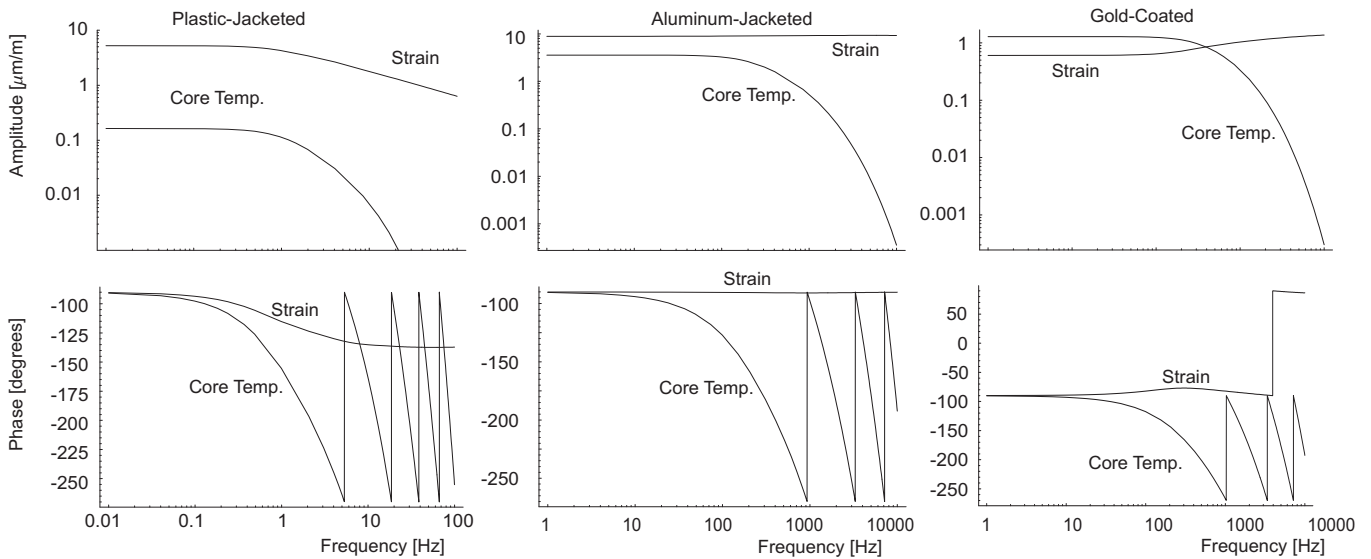,width=\textwidth}
\caption{\label{allresponse} Bode plots. Amplitude responses have been multiplied by the frequency $f$. We assume $h=2\pi$\,W/mm$^3$ heating power density (corresponding to 1.9, 1.1, and 0.25\,W, respectively, per meter of fiber length).}
\end{figure*}

The core temperature and the strain term for this fiber is shown in Fig. \ref{allresponse} (left). For better presentation in the figure, the amplitudes have been multiplied by the frequency $f$.

For frequencies below about 1\,Hz, the amplitudes of the core temperature and the strain term are proportional to $1/f$; their phases equal $-90^\circ$. This is because we neglect heat dissipation to the environment, which means that the actual thermal energy within the fiber is the integral of the heating rate over time. In practice, this is a good approximation for frequencies above about $0.1\,$Hz. At lower frequencies, the temperature of the fiber will be given by an equilibrium between heating and dissipation. Thus, the frequency response of the strain and core temperature terms will become constant below about 0.1\,Hz. The phase will then be 0$^\circ$. It is, of course, possible to incorporate dissipation into the above model. However, the heat dissipation is hard to quantify because of air currents, Moreover, it is irrelevant for the design of the fiber length feedback loop as it has significant effects only for frequencies much below the loop bandwidth. 

For frequencies above about 0.5\,Hz, the magnitude of both terms begins to drop, and the phase lags increase correspondingly. For the core temperature term, this is because fast temperature changes are attenuated on their way to the core due to the finite thermal conductivity. The core temperature term drops rapidly with increasing frequency and gains a large phase shift, which exceeds 180$^\circ$ for frequencies as low as 1\,Hz. This bandwidth is much too low for an effective removal of the path length fluctuations by a servo loop. 

Study of the strain term $\Delta L$, however, reveals a much more favorable behavior. Like the core temperature term, it starts to drop for frequencies above about 0.5\,Hz, but only very mildly, proportional to $1/f^{3/2}$. This because the strain term does not depend on the slow heat diffusion to the core alone. Rather, strain generated in the outer layers of the fiber, which are closer to the heater, is transmitted to the core at the velocity of sound. The $1/\sqrt{f}$ rolloff (in addition to the overall $1/f$ one) arises because the depth to which the temperature changes penetrate the fiber becomes thinner with increasing frequency. This mild frequency response is crucial for the performance of the servo: The $1/f^{3/2}$ rolloff at $135^\circ$ total phase shift enables a fast and stable servo loop. 

The actual frequency dependence of the strain term shown in Fig. \ref{allresponse} (left) is a consequence of the very low elastic modulus of the heater coating. A heater out of a more rigid material will cause a different, and even more interesting, response. This is made evident in the case of aluminum coated fiber, see below.

\paragraph{Aluminum-coated fiber} 

Aluminum coated fiber, where the Al coating is used as the heater, has a frequency response which is even more suitable for the present application, see Fig. \ref{allresponse} (middle). Apart from a general $1/f$ response, the core temperature term begins to roll-off at a frequency of $f_1\sim 250$\,Hz, much larger than for plastic jacket. This is because of the much larger thermal conductivity of the Al coating as compared to acryl, and because the Al coating as a heater attaches dirctly to the core.

The frequency response is again dominated by the strain term. The major difference to the plastic-jacketed fiber is that the strain term shows virtually no roll-off other than the usual $1/f$ behavior for all frequencies at which the model is valid. This is because the Al coating as a high elastic modulus, comparable to the one of the quartz glass core. As the Al layer is much thicker than the core, the stiffness of the fiber is almost completely due to the Al heater layer. Heat is generated throughout the volume of the Al coating, and there it causes strain instantaneously. This strain is propagated to the core at the high velocity of sound.  

\paragraph{Bare, gold-coated fiber}

The frequency response of fiber coated with a thin layer of gold is shown in Fig. \ref{allresponse} (right). The gold coating is used as the heater. The strain term is not dominant at low frequencies (because the stiffness of the fiber is dominated by the quartz glass core, with its low thermal expansion coefficient), but overtakes for larger frequencies. The magnitude of the strain term depicts a roll-off of slightly less than $1/f$ for frequencies between 100 and 10000\,Hz. The reason can be explained as follows: The thermal expansion coefficient of quartz glass is much lower than that of gold. Most stress is thus generated in the thin gold coating, and the resulting strain is reduced by the core due to its low thermal expansion coefficient. At low frequencies, the core also takes away a significant fraction of the thermal energy, thus reducing the amplitude of the temperature in the heater volume, thereby reducing the generated stress. At high frequencies, however, the finite thermal conductivity reduces the heat transfer to the core, so that the temperature variation is no longer reduced by heat transfer.

\paragraph{Discussion}

The phase modulation depth that can be achieved by heating is limited by the permissible temperature and hence the power dissipation of the heater. Assuming a very moderate heater power of 1\,W per meter of fiber length, we obtain a modulation depth of about 2.4\,$\mu$m/m, 8$\,\mu$m/m, and 4\,$\mu$m/m for plastic-, aluminum-, and gold-coated fiber at 1\,Hz, respectively. The higher values are obtained for the aluminum- and gold-coated fibers, where this figure drops like $1/f$ (For the plastic-jacketed fiber, it drops with $1/f^{3/2}$). 

The validity of the above model is limited at higher frequencies, since we assumed instantaneous sound propagation. Sound, however, propagates at a velocity of several 1000\,m/s. For a fiber radius of 400\,$\mu$m, this means that fiber will deviate from the above model on a time scale of the order of 0.1$\,\mu$s, or frequencies of a few MHz. Since the additional delay due to the finite velocity will increase the phase lag at high frequencies, this represents a theoretical limit on the speed of any feedback loop based on heating of a fiber. In view of the very limited modulation amplitude that can be reached at such high frequencies, however, heating of the fiber cannot take out a disturbance of any appreciable amplitude at such frequencies, even if the linearized feedback loop would theoretically be stable.

\section{Apparatus}\label{apparatus}

\subsection{Overview}

\begin{figure}
\centering
\epsfig{file=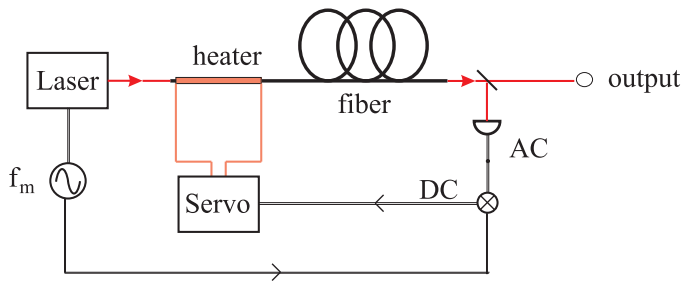,width=0.5\textwidth}
\caption{Basic setup for active length control of an optical fiber: The frequency modulated laser light
is sent through the fiber and detected by a photodetector. Demodulation with the modulation frequency yields an error signal, that is used by a servo to drive a heater in order to compensate for fluctuations in length. It is also possible to place the detector and beamsplitter to the input side of the fiber.
\label{etalon}}
\end{figure}

The basic setup is shown in figure \ref{etalon}. For reading out the instantaneous length of the fiber, we use the method developed by Pound, Drever, and Hall \cite{PDH} for reading out the resonance frequency of a resonant cavity. Indeed, the fiber with its about 4\% parasitic reflection at each end resembles a Fabry-Perot etalon, although one with a quite low finesse. (However, other methods for reading out the actual fiber length could be used, like the one employed by Ma {\em et al.}) Frequency modulated Nd:YAG laser light at 1064\,nm is sent
through the fiber. The modulation at a frequency of $f_m=500$\,kHz with a modulation index of about 1 (however, these parameters are not critical for the performance of the active length control) is generated by modulating the strain of the Nd:YAG monolithic ring cavity with a piezo. From the output of the fiber, we split off a sample and detect its intensity. Alternatively, the detector can be placed at the input of the fiber. This makes the setup extremely simple, as no additional components are required at the remote end of the fiber.

Due to the frequency modulation (FM), the laser radiation acquires additional Fourier components at frequencies $f_l\pm f_m$ (``sidebands"), where $f_l$ is the laser frequency. For pure FM, the phase relationship between these components is such that the amplitude of the signal is unaffected. The resonant and dispersive properties of the fiber alter these phase relationships. In general, this converts the pure frequency modulation into amplitude modulation (AM) components at multiples of $f_m$. When the laser and
the resonance frequency coincide, the detected component of the AM vanishes; if they are detuned, however, a nonzero AM is found, whose amplitude indicates the magnitude and whose phase relative to the FM indicates the direction of the detuning. The detector signal is multiplied with a local oscillator (LO) signal at $f_m$ in a double-balanced mixer (DBM), and the output of the DBM is low-pass filtered to suppress the modulation frequency. With the correct LO phase $\phi$, a error signal is thus generated that indicates the deviation of the optical path length of the fiber relative to a multiple of half the laser wavelength. 

This signal serves as input to a proportional-integral servo, which controls the fiber length by heating a part of the fiber. For determining the gain and time constants of the servo, we use the results of our above study of the the dynamic response of the fiber (Fig. \ref{allresponse}). Since a servo loop is stable for a $1/f^{3/2}$ rolloff with a corresponding phase shift of 135 degrees, it can operate throughout the frequency range discussed in the above model. 

The most suitable fiber would be the aluminum-jacketed or gold-coated ones (Tab. \ref{fibers}). The most undesirable one, from the point of view of high servo speed, is plastic-jacketed fiber, because the plastic jacket is thick and has poor thermal conductivity. The results of the previous section, however, show that even in this case loop bandwidths of several 100\,Hz can be achieved.

To demonstrate that even in this case a satisfactory performance can be obtained, we used a 5\,m long polarization maintaining fiber with plastic jacket (Thor labs type FS-PM-4511). We tried two different methods to attach a heater layer to it: A manganine wire wrapped around the fiber did not work because of the bad thermal contact between the wire and the fiber. Much better contact was obtained by coating the fiber with electrically conductive silver varnish on a length of $30$ cm. 

\begin{figure}[t]
\centering
\epsfig{file=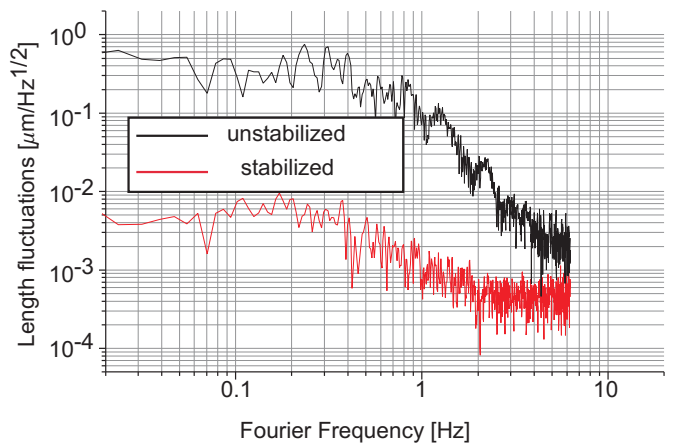,width=0.49\textwidth}
\caption{Error signal with and without stabilization. \label{laenge}}
\end{figure}

To examine the length stability of the fiber, we monitor the error signal with the servo loop open and closed. Without stabilization, the fluctuations reach up to $0.5 \mu$m/$\sqrt{\rm{Hz}}$ at 1 Hz and increase proportionally to $1/f^2$ for lower frequencies. Fig. \ref{laenge} does only show this for frequencies above 1\,Hz, because the length readout provided by the Pound-Drever Hall method is restricted to a range of $\lambda/2$. Applying active length stabilization, the fluctuations are reduced below 8\,nm/$\sqrt{\rm Hz}$ for frequencies below about 1\,Hz, and below 1\,nm$\sqrt{\rm Hz}$ above 1\,Hz. The fluctuations integrated over the frequency range of Fig. \ref{laenge} are about 10\,nm, or $\lambda / 100$. This measurement demonstrates the performance of fiber heating as an actuator to reduce any measured fiber length fluctuations. With negligible errors in the readout of the fiber length by the PDH method in the servo gain limited regime, this is also the level of residual fiber length fluctuations. At this point, however, it should be noted that this in-loop data does not reveal possible errors made in the readout of the fiber length by the PDH method. 

Given the simplicity of our setup, these results compare well to those due to Ma {\em et al.} \cite{Ma}: For  25\,m of fiber, within a frequency band of $\sim 0.01-2\,$kHz, they suppresed the phase noise to a level of -60\,dBc/Hz, corresponding to residual length fluctuations below $\sim 1\,$nm/$\sqrt{\rm Hz}$. The noise is increasing to $1\,\mu$m/$\sqrt{\rm Hz}$ at $0.1\,$Hz, but this is a very conservative value, as Ma {\em et al.} infer it directly from a beat measurement between the incident and the transmitted light. Although the bandwidth of our lock ($\sim200$\,Hz) is lower than in the previous work, our setup achieves a similar performance for low and medium frequencies.

In many cases, the frequency modulation used for
reading out the path length will not interfere with the intended
application of the fiber's output signal. The sidebands form
separate components in frequency space; in a frequency comparison
by taking a beat note with another optical signal, for example,
they can be removed by electronic filtering. Its effects
can be reduced by using a low modulation index and by choosing a suitable modulation
frequency. 

The coating with silver varnish turned out to be unsuitable for a long term usage. Occasionally, it breaks at its weakest point, which makes further length controlling impossible. Although the defect can easily be repaired and the fiber itself does not suffer any damage, a reliable long-term performance requires a more stable solution, like using of aluminum-jacketed fiber (Tab. \ref{fibers}).

\section{Application within a cryogenic resonator frequency
standard}\label{coresec}

\begin{figure}
\centering
\epsfig{file=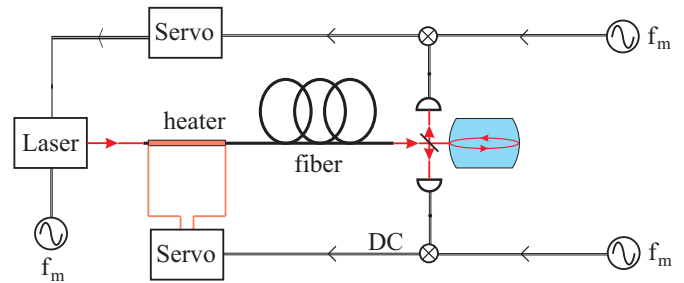,width=0.49\textwidth}
\caption{Fiber length stabilization in a CORE frequency standard
\label{core}}
\end{figure}

We applied the method in an optical frequency standards based on a Nd:YAG laser stabilized to a cryogenic optical resonator (CORE), see figure \ref{core}. This consists of an oscillator, a monolithic, diode-pumped neodymium-YAG laser, frequency stabilized to a CORE located inside a liquid helium cryostat. The frequency stabilization uses the Pound-Drever-Hall method using a photodetector that detects the light that is reflected from the core (Fig., \ref{core}, above). In previous work, this system was shown to reach a Hz-level stability on a timescale of seconds. 

\begin{figure}[t]
\centering
\epsfig{file=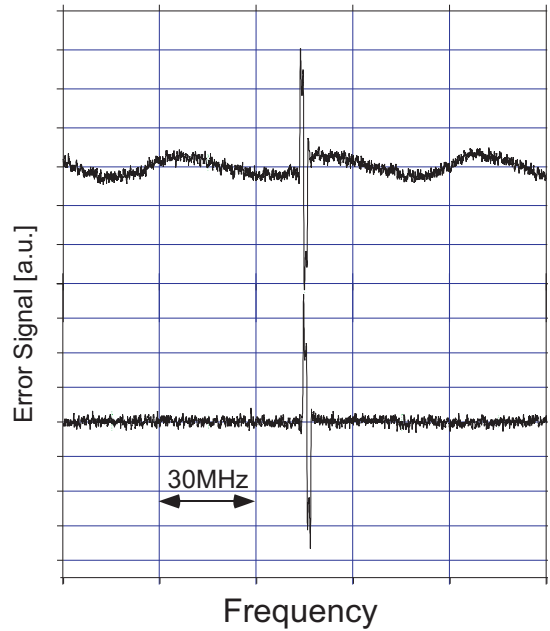,width=0.4\textwidth}
\caption{Error signals without and with active fiber length stabilization.
\label{etalonsig}}
\end{figure}

An optical fiber is used for coupling the laser light to the cavity (Fig. \ref{core}). While this makes the system more stable on timescales of hours \cite{BraxmaierKT,CPEM}, on short timescales, however, the frequency stability with fiber coupling is reduced due to path length fluctuations. These introduce phase fluctuations as described above, as they shift the laser frequency relative to the CORE resonance. Additionally, since the fiber with parasitic reflections at its ends acts like a Fabry-Perot etalon, it adds parasitic signal components into the Pound-Drever-Hall error signal used for feeding back to the laser frequency. These change from maximum to minimum for a $\lambda/4$ length change of the fiber and cause an unwanted fluctuation of the baseline of the error signal (Fig. \ref{etalonsig} (above)), which have been found to lead to fluctuations of the Nd:YAG laser frequency of 8\,Hz amplitude.

To remove both the Doppler effect and this modulation, we lock the path length of the fiber to the Nd:YAG laser's wavelength, which is itself locked to the CORE. A photodetector located inside the cryostat (Fig. \ref{core}, below) is used to read out the fiber length fluctuations. Its signal is down converted with the laser modulation frequency to generate an error signal for reading out the fiber length (separate sources for $f_m$, which are synchronized to each other, allow an easy adjustment of the phase of the local osciallator signals). It is fed back to the heater via a proportional-integral controller. The fiber and the heater are of the same type as described above.

Figure\,\ref{etalonsig} illustrates the effect of fiber-length stabilization. Above is shown the error signal without active fiber length stabilization, measured by scanning the laser frequency over a range of about 150\,MHz. Note the baseline fluctuations caused by the parasitic etalon due to the fiber with a free spectral range of about 50\,MHz. Active fiber length control tracks the fiber length according to the actual laser wavelength and removes these effects (Fig. \ref{etalonsig} (below)). For the 150\,MHz laser frequency sweep, the fiber length changes by about 2.5$\,\mu$m to track the fiber length to the actual laser wavelength. Thus, the fiber is always operated with maximum transmission, where the front and the back reflex cancel. A straight baseline free of etalon fringes seen in Fig. \ref{etalonsig} (above) and a constant signal amplitude at maximum value are obtained. The speed of the servo loop made it possible to track the laser wavelength even while it is scanned at a rate of about 10\,Hz.

\section{Summary and Outlook}

We have presented a method to reduce fluctuations of the optical path length of optical fibers to below 10\,nm by controlled heating of a part of the fiber. A theoretical study of the response of the optical path length to heat dissipated in the bulk of a heater layer indicates that servo bandwidths in the kHz range, or indeed even the low MHz range, can be obtained. This high speed is possible because strain is instantaneously generated at the heated surface. It then propagates to the fiber core at the speed of sound, which is a much faster process than heat diffusion. Bare fiber with a thin gold coating for resistive heating, or aluminum jacketed one are especially suitable.  

We also present a first experimental demonstration of the method. It uses a conductive coating made of silver varnish on a plastic-jacketed fiber. We used this method in the context of an optical frequency standard based on a cryogenic optical resonator. 

Compared to previous work, our method has the advantage of directly removing the path length fluctuations, rather than compensating for them via an acousto-optic modulator (AOM). Thus, it can reduce the phase noise induced by such fluctuations for any signals, even those for which AOMs are not suitable. An example for such a signal are broadband optical signals like those generated by optical frequency comb generators. It also avoids the loss due to the finite deflection efficiency of an AOM. A modulator may be necessary to provide the phase modulation used for measuring the actual fiber length. However, as in the example above, such a modulation may already be present in a particular setup. Else, it can be generated by an electro-optical modulator, which has much lower insertion loss than an AOM, and whose operation does not depend on the laser wavelength as the one of an AOM. In principle, the modulation only needs to be applied for one reference beam that provides for the fiber length measurement. Thus, this modulation is no obstacle in applications with a large wavelength range. 

The simple realization of the active stabilization leaves plenty of space for further developments. Use of aluminum-jacketed fiber can enhance the reliability and also increase significantly the useful lock bandwidth. This leads to better stability of the path length. Another approach is bare fiber coated with a gold layer, which leads to even higher speed. Also, one can study the influence of (and optimize) the environment the fiber is in. Presumably, by thermal and acoustical shielding, a substantial improvement of the noise can be reached, at least for certain noise frequencies. In analogy to the method of Ma {\em et al.}, however, we expect that our method surpasses the stability of such passive stabilization (see, for example, \cite{deBeauvoir}) by 20-30dB \cite{Ma,Ye2003}.

We expect that the method can soon be routinely applied in many applications mentioned in the introduction, that require the transmission of an optical signal through fiber with exceptional stability. To give an example, consider the 3.5\,km long fiber in Ref. \cite{Ye2003}. The feedback bandwidth used in this work was about 10\,kHz, which can be attained, for example, with the Al- or Au-coated fibers as described above. Thus, we have a sufficiently fast actuator to obtain a similar degree of suppression of perturbations. A similar level of accuracy should thus be obtained. To achieve sufficient dynamic range, it would be necassary to equip no more than about 1/10 of the total fiber length with the heater (assuming that the dynamic range of the heater is more than 10 times larger than the variation of the external temperature). However, the easiest way would probably be to obtain a commercial Al-jacketed fiber and to control the whole path. The question is, which method is easier to implement? Obviously, the old method allows to use any kind of fiber. Ours, on the other hand, has advantages if the loss associated with the AOM is critical or if a broad range of laser frequencies is to be transmitted at the same time (like the simultaneous transmission of a laser frequency along with its first harmonic or even a femtosecond frequency comb). The phase shift of those signals could not be compensated for using a single AOM.   

\section{Acknowledgements}

We thank Stephan Eggert for building excellent electronic circuits. H.M. whishes to thank S. Chu and S.-w. Chiow for discussions and the Alexander von Humboldt-Stiftung for support.

%\begin{figure}[t]
%\centering \epsfig{file=schema.eps,width=0.7\textwidth}
%\caption[]
%{Schematischer Aufbau der Erfindung: 1: Frequenzquelle,
%2: Regler, 3: Stellelement, 4: \"{U}bertragungstrecke, 5: Information, 6,7: Fehlersignale, 8:
%Stellsignal.}

%\label{Aufbau}
%\end{figure}

\end{document}